\documentclass[11pt,a4paper,onecolumn,aps,nofootinbib,showpacs,floatfix]{revtex4}

\usepackage[cp1250]{inputenc}
\usepackage{graphicx} 
\usepackage{epstopdf} 
\usepackage{multirow}

\begin{document}

\title{Intra-day variability of the stock market activity versus stationarity \\of the financial time series}
\author{T. Gubiec}
\email{tomasz.gubiec@fuw.edu.pl}
\author{M. Wili{\'n}ski}
\email{mateusz.wilinski@fuw.edu.pl}
\affiliation{Institute of Experimental Physics,\\ Faculty of Physics, University of Warsaw  \\ Pasteura 5, PL-02093 Warsaw, Poland}

\begin{abstract}
We describe the impact of the intra-day activity pattern on the autocorrelation function estimator. 
We obtain an exact formula relating estimators of the autocorrelation functions of non-stationary process to its stationary counterpart. 
Hence, we proved that the day seasonality of inter-transaction times extends the memory of as well the process itself as its absolute value.
That is, both processes relaxation to zero is longer.
\end{abstract}


\maketitle

\section{Introduction}\label{section:intro}

Time series of logarithmic price returns are commonly used in a broad range of financial analysis. 
Recently, the intra-day type of this data focused a  particular interest. 
While using many different types of estimators and models, it is often assumed, directly or indirectly, that underlying processes are stationary. 
Unfortunately, it is not the case for financial time series, even for their logarithmic returns. 
There are at least few well known reasons against financial data stationarity. One of them is \textit{volatility clustering}, a positive autocorrelation observed for different measures of volatility. 
This effect was described among others by Tsay \cite{Ts}, Cont \cite{Co} and Guillaume \textit{et al.} \cite{Gu}. 
Complementary aspect, more closely associated to this work, are different types of seasonalities, that can be seen in various time scales. 
In a year scale we have for example "Santa Claus rally" which is associated with rise of stock prices in December. 
We can also observe seasonalities in month and week scale but a major example is so called 'lunch effect' or intra-day pattern which refers to day trading and is characterized by high volatility and short inter-trade times right at the beginning of the session and just before closing of the quotations, and significantly lower volatility in the middle of the day. 
Intra-day changes of the activity on stock market is a well-known empirical fact observed all around the world on different types of market. More details can be found in Hasbruck \cite{Ha}, Chan \textit{et al.} \cite{Ch}, Gençay \textit{et al.} \cite{Ge}, Admati and Pfleiderer \cite{Ad}.

\begin{figure}
	\begin{center}
     \includegraphics[width=3in]{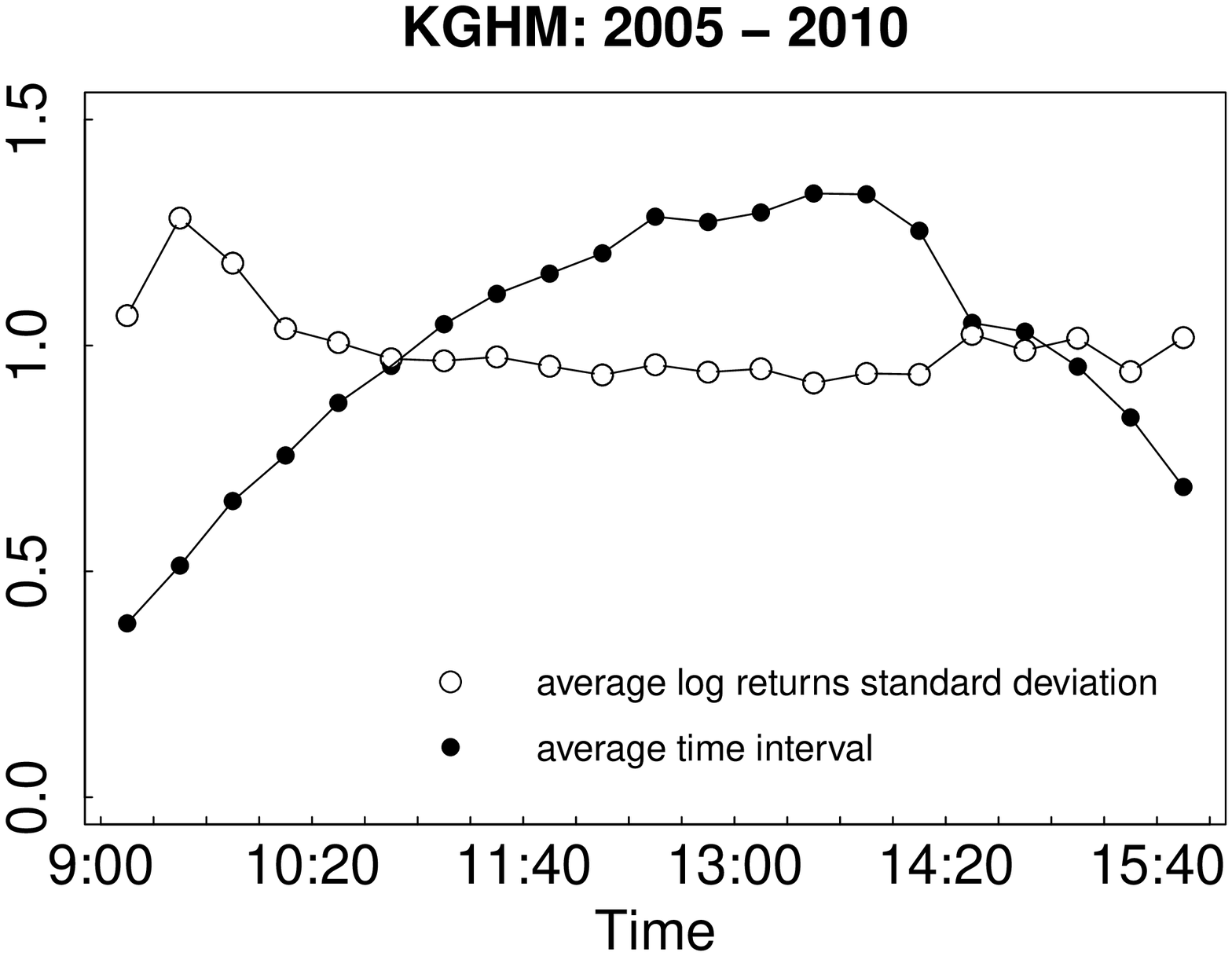}
     \includegraphics[width=3in]{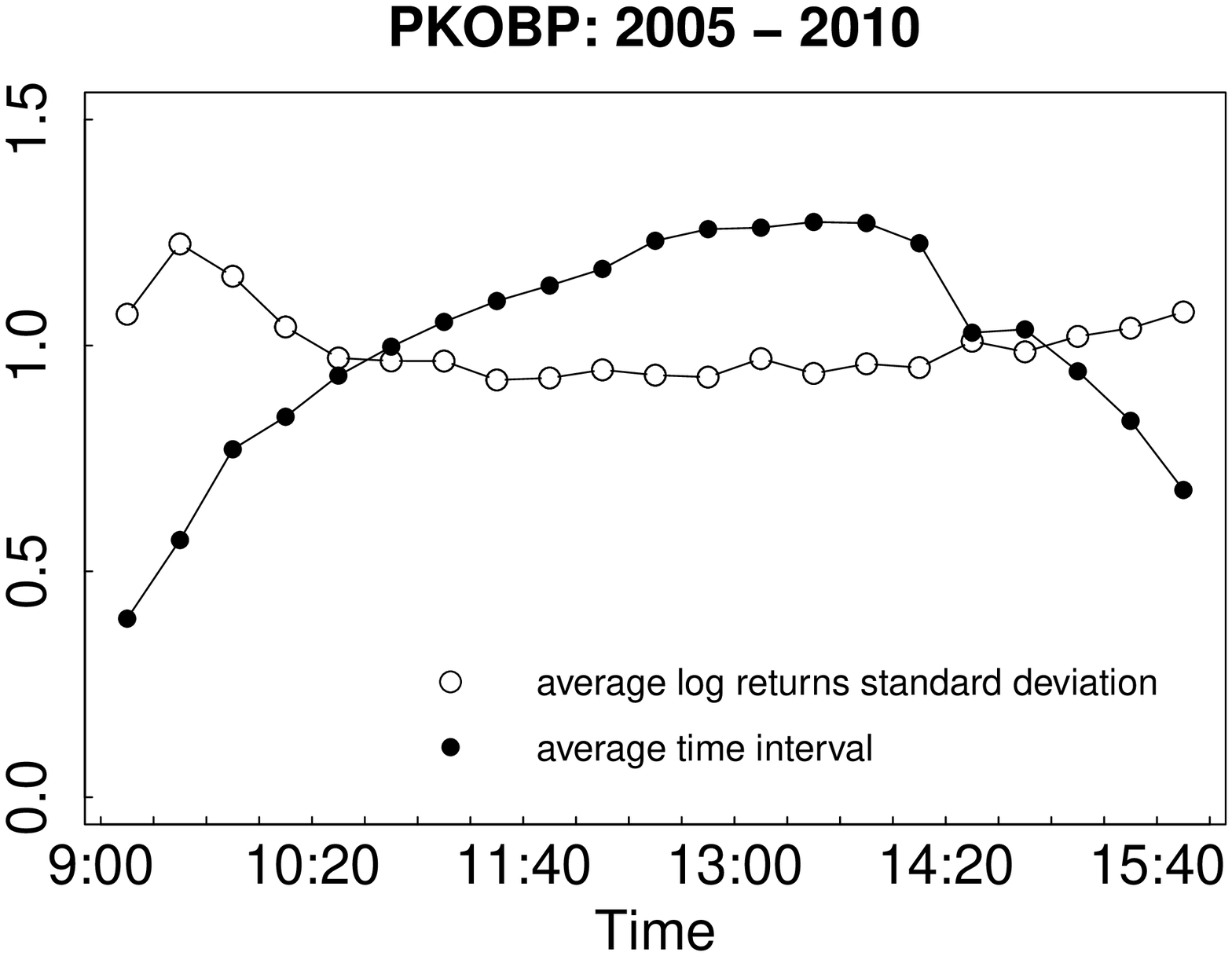}
	\end{center}
	\caption{Intra-day seasonality in both standard deviation and inter-transaction times calculated in 20 minutes time intervals (both averaged over days and then divided by their mean, in order to make them comparable), shown for two companies listed at Warsaw Stock Exchange. In both cases, we observe that average time interval doubled or even tripled its value, whereas average standard deviation of price returns changed not more then several dozen percents. Results on the left were obtained by using all transactions for KGHM in years 2005-2010. The second plot concerns PKOBP in the same period. It should be pointed out that from 2005 to 2010 WSE was open for 7 hours.}
	\label{pic:season}
\end{figure}

Previously, the research focused  mainly on the variability and seasonality of volatility. 
Part of it measured theirs impacts and removed their effect from the data \cite{Am}. 
Such an approach doesn't affect time intervals between trades at all, while the range of average inter-trade time is greater than the range of standard deviation of returns (see Fig. \ref{pic:season} for details).
Other commonly used approach to this problem is a restriction of the data to the so called event time representation.
However, by using this method, we would not be able to draw any conclusion on the time dependence. 

The aim of this work, is to find how the intra-day seasonality, observed in the inter-transactions time intervals, affects the autocorrelation of the time series. 
We propose the systematic analytical approach to the time transformation, which eliminates this seasonality from analyzed process.
Below, we focus on the impact of the inter-transaction time seasonality on the estimators of the autocorrelation functions. 
Although, taking into account the variability of standard deviation of price returns is possible, in this paper we focus on the impact of the former effect.

\section{Analysis}\label{section:analysis}

The intra-day pattern can be visualized by the number of trades executed in subsequent periods of time during the day. 
Equivalently, this can be shown by the average time gaps between transactions in these periods of time, as it is shown in Fig. \ref{pic:season}.
The average is taken, herein, over the statistical ensemble of days and inside the chosen period of time. 

We propose two different functions to describe how average interval between trades changes over time. 
Then, we use these functions to  transform price returns process in order to dispose its seasonality. 
Hence, we find a relationship between autocorrelation functions of the process before and after the transformation.

\subsection{Analytical form of day seasonality}\label{section:season}

We can assume the relationship between a clock time and average inter-trade interval in a functional form. 
We introduce two different functions to describe this relationship:
\begin{equation}
\label{eq:theta}
\theta^{(1)} (t) = a(t - t_1)(t - t_2) \quad \mbox{and} \quad \theta^{(2)} (t) = \frac{1}{a((t - p)^2 + q)}.
\end{equation}
The first of them is a quadratic function driven by parameters  $a$, $t_1$ and $t_2$. 
The second function is a rational one driven by parameters $a$, $p$ and $q$.

Let us assume that $Y(t)$ (representing the quantity which autocorrelation we analyze) is a process with a seasonality described above. 
This process represents e.g. returns and their absolute values. 
Furthermore, we assume that there exist an underlying process $X(\tau)$, which is ergodic and therefore stationary (without seasonalities). 
By term 'underlying' we mean a direct relationship between $X(\tau)$ and $Y(t)$, that is described by relation:
\begin{equation}
\label{eq:base}
Y(t) = g(t)X(\tau(t)),
\end{equation}
where $\tau=\tau(t)$ is responsible for presence of seasonality in apparent process $Y(t)$. 
In our case $t \in [0,T]$, where $t = 0$ is the beginning of trading day and $t = T$ is the end of quotations. 
We require our transformation $\tau(t)$ to preserve the day length and to change the time intervals only, which leads to $\tau(t): [0,T] \rightarrow [0,T]$, where $\tau(0) = 0$ and $\tau(T) = T$. 
Also, the number of events (transactions) is the same for both processes. 
Hence, the daily average time intervals between transactions are equal in time space $t$ and $\tau$, i.e. $\langle t \rangle = \langle \tau \rangle$.

\begin{figure}
	\begin{center}
     \includegraphics[width=3in]{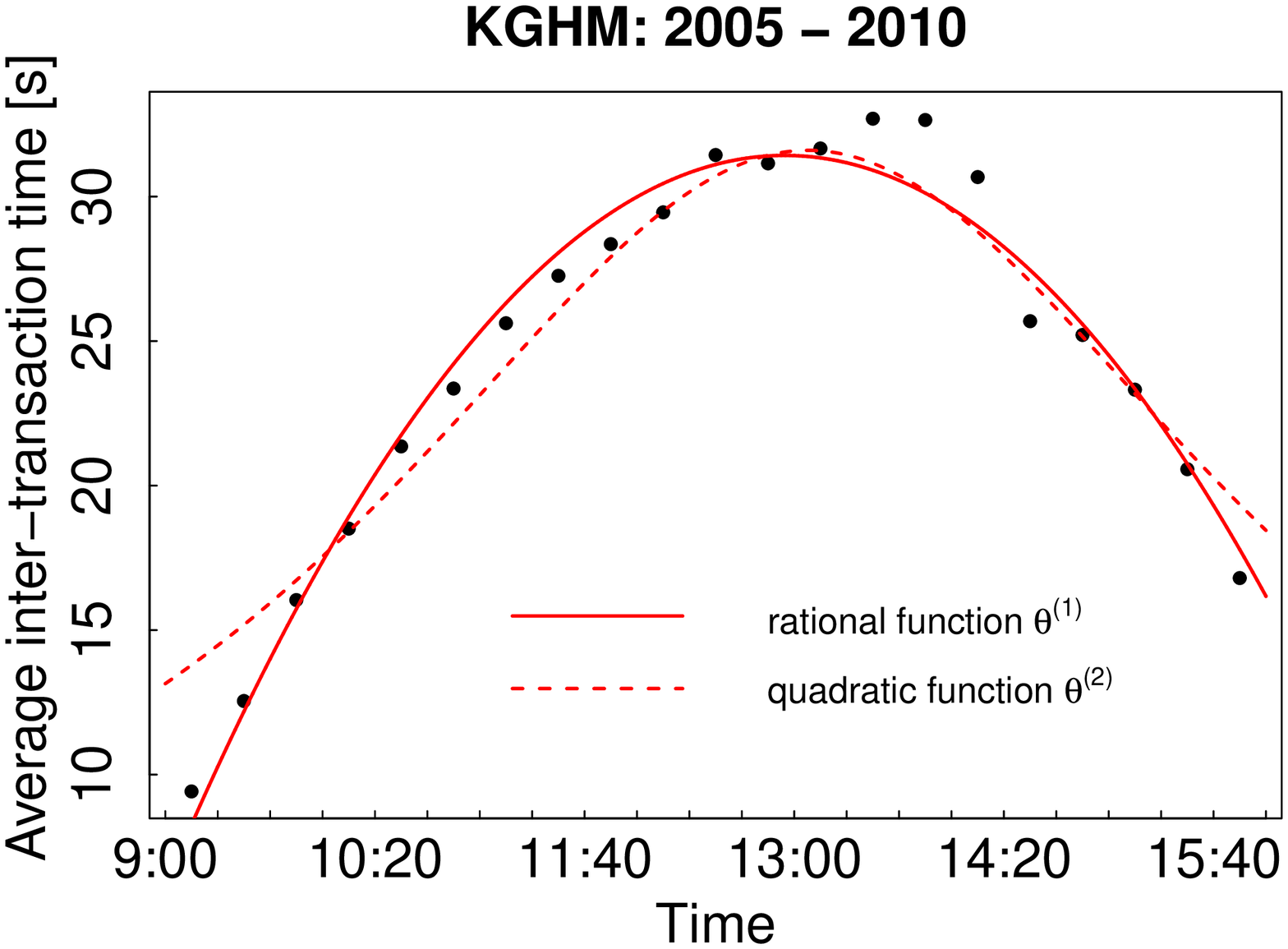}
     \includegraphics[width=3in]{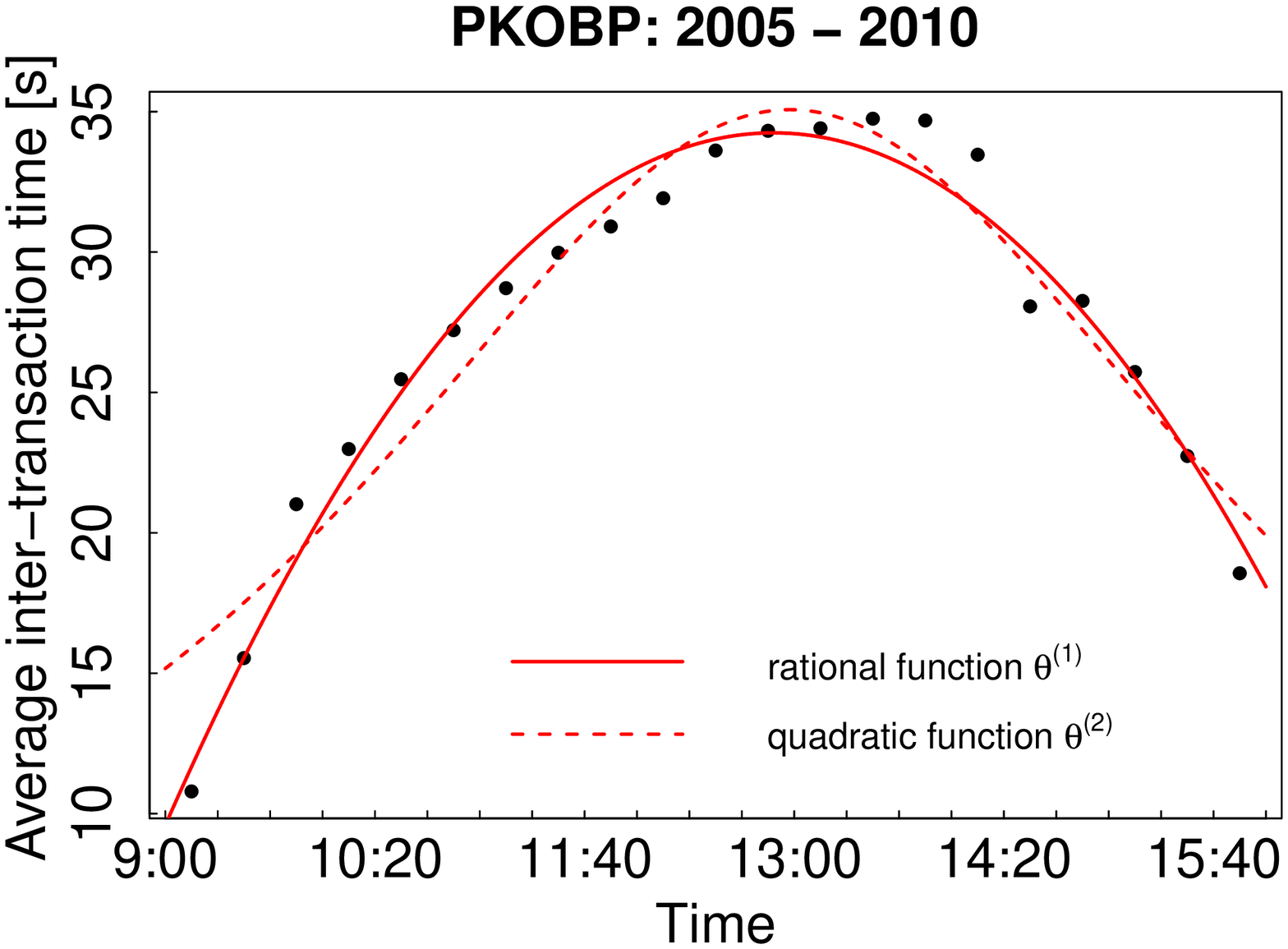}
	\end{center}
	\caption{Graph of intra-day pattern, for instance, for  KGHM and PKOBP (black circles) with imposed functions $\theta^{(1)}$ and $\theta^{(2)}$ (solid and dashed curves, respectively).}
	\label{pic:theta}
\end{figure}

Function $\tau(t)$ is strictly related to the time dependent average inter-transaction time, that is to function $\theta (t)$. 
The average inter-trade interval for $X(\tau)$ should be constant and equal $\langle \tau \rangle$. 
Therefore, we stretch the time-line $t$ when average intervals are small and compress it when these intervals are too long. 
As $\theta (t)$ is the function that determines average trade interval for process $Y(t)$, we can write:
\begin{equation}
\label{eq:base2}
\frac{dt}{\theta (t)} = \frac{d \tau}{\langle \tau \rangle},
\end{equation}
where, as said above, $\langle \tau \rangle = \langle t \rangle$ is an average inter-trade interval for both $Y(t)$ and $X(\tau)$. 
By integrating both sides of this equation we obtain the relationship between $t$ and $\tau$:
\begin{equation}
\label{eq:f}
\tau(t) = \langle t \rangle \int^{t}_{0} \frac{1}{\theta (s)}ds.
\end{equation}
When analyzing empirical data, day length $T$ and the number of transactions are known and we can easily calculate $\langle t \rangle$. 
Substituting $\theta^{(1)}$ and $\theta^{(2)}$ into Eq. (\ref{eq:f}) we get
\begin{equation}
\label{eq:av_t}
\langle t \rangle^{(1)} = \frac{a T (t_2 - t_1)}{\ln{\left( \frac{(T - t_2)t_1}{(T - t_1)t_2} \right)}} \quad \mbox{and} \quad \langle t \rangle^{(1)} = \frac{1}{a \left( \frac{T^2}{3} - p T + p^2 + q \right)}.
\end{equation}
In both cases the constraints above allow us to reduce the number of parameters in $\theta$s to two.
Fig. \ref{pic:theta} shows $\theta^{(1)}(t)$ and $\theta^{(2)}(t)$ fitted both to KGHM and PKOBP. 
The only fitted parameters are $t_1$, $t_2$ for $\theta^{(1)}$ and $p$, $q$ for $\theta^{(2)}$, as parameter $a$ is determined by using Eq. (\ref{eq:av_t}) and the empirical average time.

\begin{table}
	\setlength{\tabcolsep}{15pt}
	\centering{
	\caption{Parameters fitted to tick data from http://bossa.pl/.}
	\begin{tabular}{cc|cc}
		{\bf Variability function $\theta$} & {\bf Parameters} & {\bf KGHM: 2005-2010} & {\bf PKOBP: 2005-2010} \\
		\hline
		\multirow{2}{*}{\bf Quadratic} & {$t_1$} & {-1640.65} & {-2485.34} \\
		 & {$t_2$} & {29999.47} & {30336.39} \\
		\multirow{2}{*}{\bf Rational} & {$p$} & {14715.08} & {14301.01} \\
		 & {$q$} & {1.54 $\cdot 10^8$} & {1.56 $\cdot 10^8$} \\
		\hline
		 \multicolumn{2}{c|}{$\langle t \rangle$} & {24.465} & {27.292} \\
		 \multicolumn{2}{c|}{No transactions} & {1624438} & {1432408} \\
		\hline
	\end{tabular}
	\label{tab:param}
	}
\end{table}


\subsection{Estimator of autocorrelation function}\label{section:corr}

When analyzing autocorrelation of empirical data, it is usually estimated with moving average. 
This estimation, in terms of the stochastic process $Y(t)$, can be described in an integral form as follows:
\begin{equation}
\label{eq:cor1}
C_Y ( \Delta t ) = \frac{1}{T - \Delta t} \int_0^{T - \Delta t} \Big( \langle Y(t + \Delta t ) Y ( t ) \rangle - \langle Y(t + \Delta t ) \rangle \langle Y ( t ) \rangle \Big) dt,
\end{equation}
where $C_Y$ is an estimator of autocorrelation function of process $Y$ at fixed lag $\Delta t$. 
Brackets $\langle \cdot \rangle$ represent an average over statistical ensemble. 
Naturally, if the process $Y(t)$ is ergodic, the estimator (\ref{eq:cor1}) converges to the real autocorrelation function of the process $Y(t)$. 
By applying Eq. (\ref{eq:base}) and using stationarity of $X(\tau)$ (its autocorrelation is a time invariant quantity), we get:
\begin{equation}
\label{eq:cor2}
C_Y ( \Delta t ) = \frac{1}{T - \Delta t} \int_0^{T - \Delta t} g (t + \Delta t) g (t) C_X (f(t + \Delta t) - f(t)) dt.
\end{equation}
It is convenient to denote $\Delta \tau (t, \Delta t) = \tau(t + \Delta t) - \tau(t)$ and to have the above integral over $d \Delta \tau$, instead of $dt$. 
By substitution $t \rightarrow t(\Delta \tau, \Delta t)$ we get 
\begin{equation}
\label{eq:cor3}
C_Y ( \Delta t ) = \int_{\Delta \tau_{min}}^{\Delta \tau_{max}} \sum_i W_i (\Delta \tau, \Delta t) C_X (\Delta \tau) d \Delta \tau,
\end{equation}
where $\Delta \tau_{min}$ and $\Delta \tau_{max}$ are the integration limits in $\Delta \tau$ space, whereas:
\begin{equation}
\label{eq:weight}
W_i (\Delta \tau, \Delta t) = \frac{1}{T-\Delta t} g(t_i(\Delta \tau, \Delta t) + \Delta t) g(t_i(\Delta \tau, \Delta t)) \left| \frac{d t_i}{d \Delta \tau} \right|.
\end{equation}
Index $i$ corresponds to different subsections of $[0,T]$, where $\Delta \tau (t)$ is injective and therefore invertible. 
Now, we can consider the estimator of autocorrelation of non-stationary process $Y$ as weighted mean of autocorrelation of stationary underlying process $X$. Furthermore, having the normalization condition:
\begin{equation}
\label{eq:prob}
\int_{\Delta \tau_{min}}^{\Delta \tau_{max}} \sum_i W_i (\Delta \tau, \Delta t) d \Delta \tau = 1,
\end{equation}
we can write $C_Y = \langle C_X \rangle_{\rho_{\Delta t}}$, where probability distribution function $\rho_{\Delta t} (\Delta \tau) \stackrel{\mbox{def.}}{=} \sum_i W_i (\Delta \tau, \Delta t)$. 
By means of this pdf way we are able to express the non-stationary estimator by the stationary one by probabilistic approach.
As we focus only on the impact of inter-transaction time variability on the autocorrelation estimators, we assume further in this text $g(t) = 1$. 

\subsection{Seasonality impact}\label{section:imp}

\begin{figure}
	\begin{center}
     \includegraphics[width=3in]{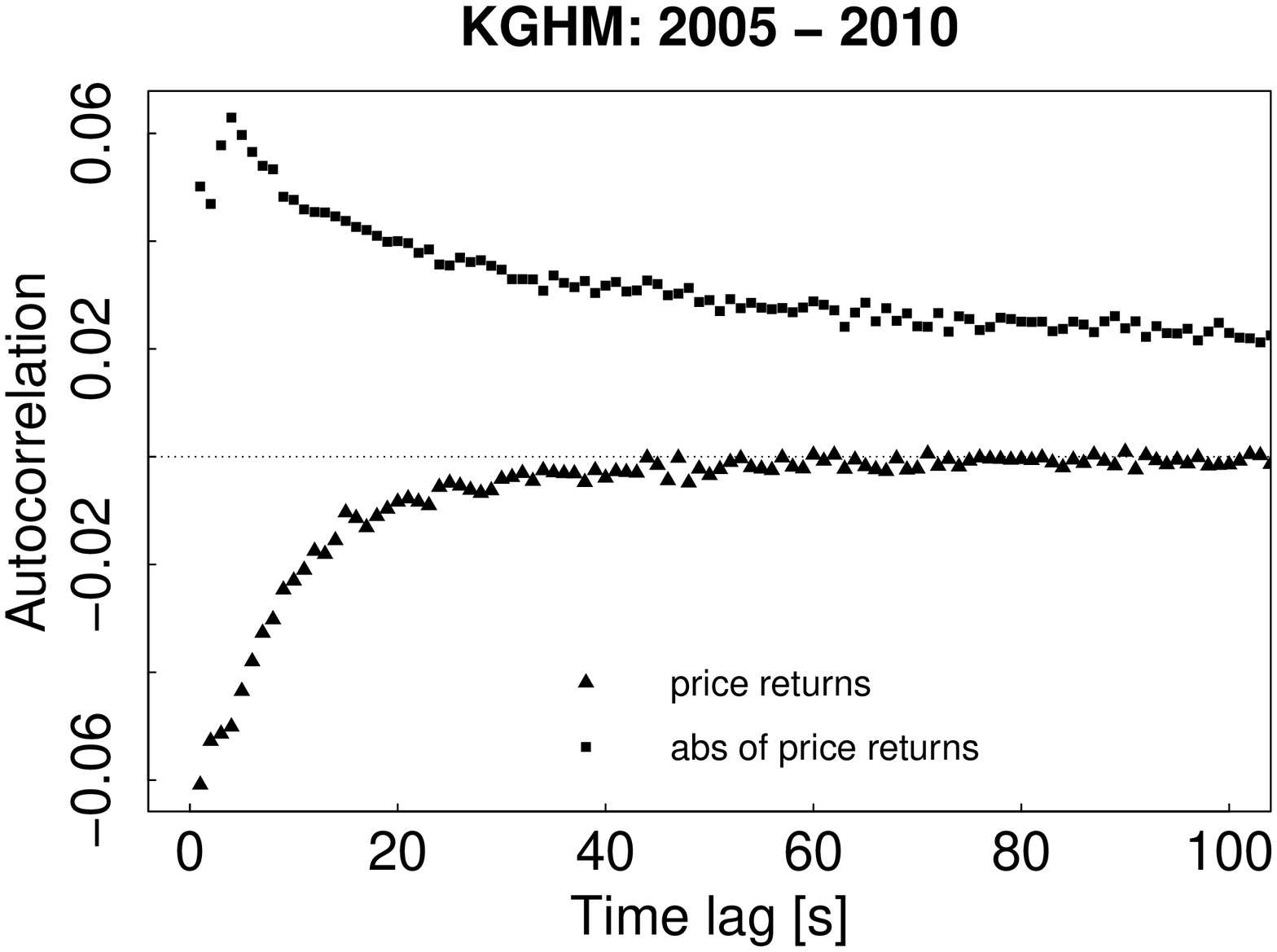}
     \includegraphics[width=3in]{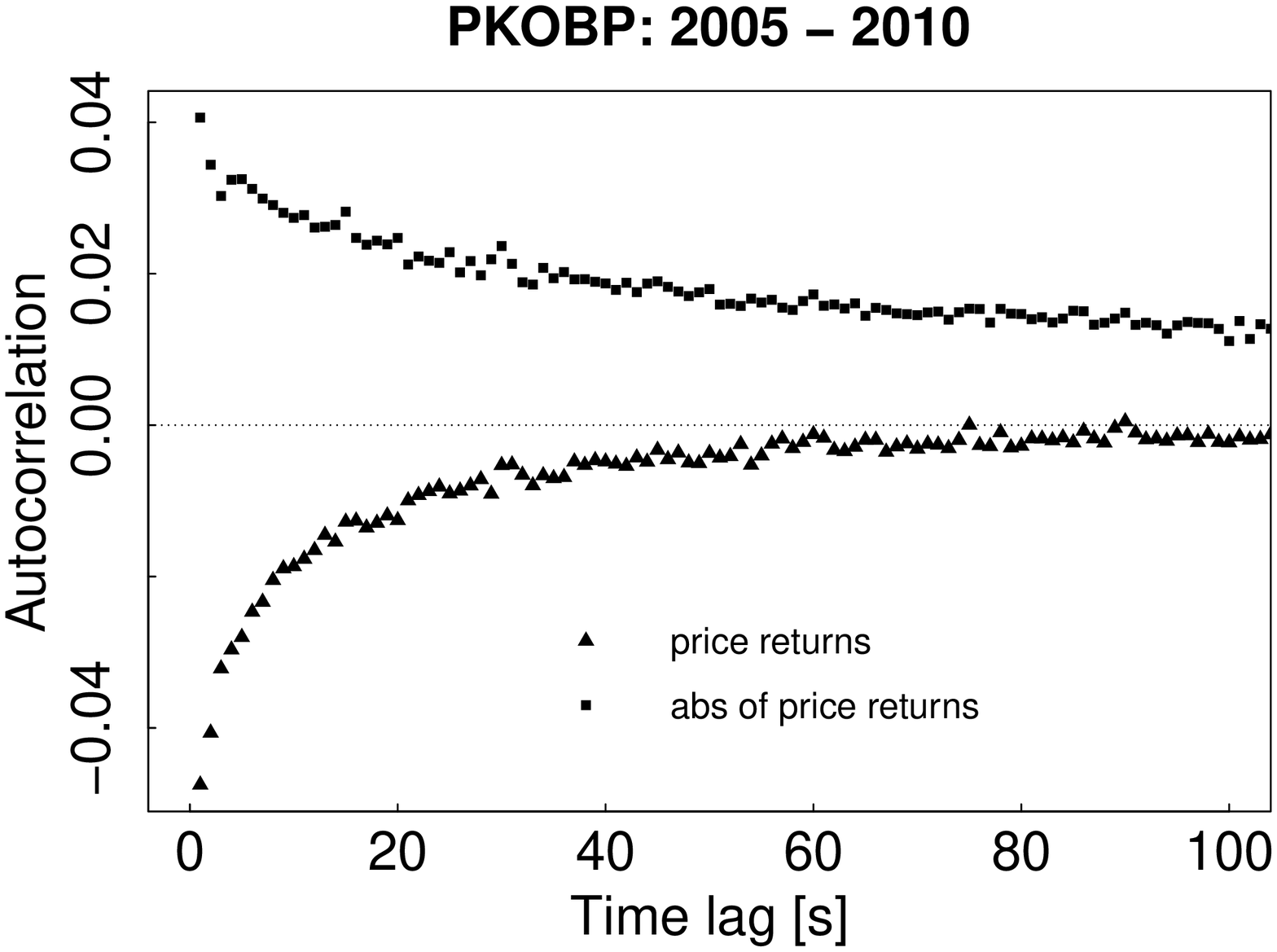}
	\end{center}
	\caption{Estimators of velocities autocorrelations of price returns and their absolute values, for KGHM and PKOBP in years 2005-2010.}
	\label{pic:autocor}
\end{figure}

In the previous section, we found the relation between estimators of the autocorrelation functions of the processes $Y$ and $X$, and their absolute values. 
The question arises: how does seasonality quantitatively affect observed estimators of autocorrelation? 
Let us start with some basic characteristics of autocorrelations observed in financial markets.
Fig. \ref{pic:autocor} shows autocorrelations of velocity estimators obtained for KGHM and PKOBP from 2005 to 2010.
In order to obtain these quantities, we use the slotting method \cite{Gub}. 
Apparently, the autocorrelation function is negative (except of $\Delta t = 0$, where it equals 1), increasing and concave for price returns but positive, decreasing and convex for price returns absolute values.

Furthermore, we use Jensen's inequality, which holds for every random variable $Z$ and any concave function $f$:
\begin{equation}
\langle f(Z) \rangle \leq f(\langle Z \rangle).
\end{equation}
For the convex function $f$ the inequality is opposite. 
As in our case $f = C_X$ and the probability distribution is $\rho_{\Delta t}$, we obtain:
\begin{equation}
\label{eq:ineq1}
C_Y ( \Delta t ) = \int_{\Delta \tau_{min}}^{\Delta \tau_{max}} \sum_i W_i (\Delta \tau, \Delta t) C_X (\Delta \tau) d \Delta \tau \leq C_X \left(\sum_i \int_{\Delta \tau_{min}}^{\Delta \tau_{max}} \Delta \tau W_i (\Delta \tau, \Delta t) d \Delta \tau \right).
\end{equation} 

\begin{figure}
	\begin{center}
     \includegraphics[width=3in]{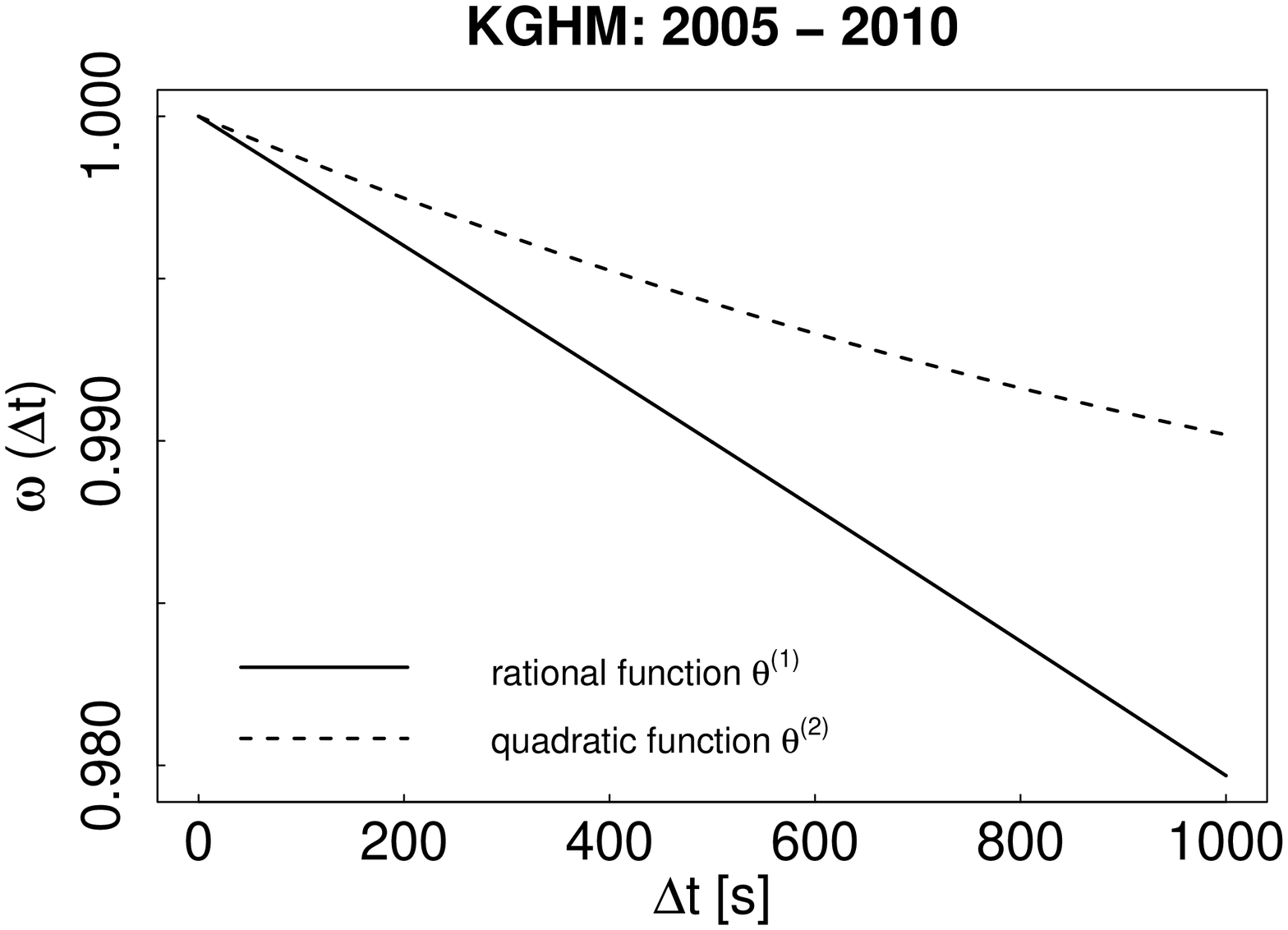}
     \includegraphics[width=3in]{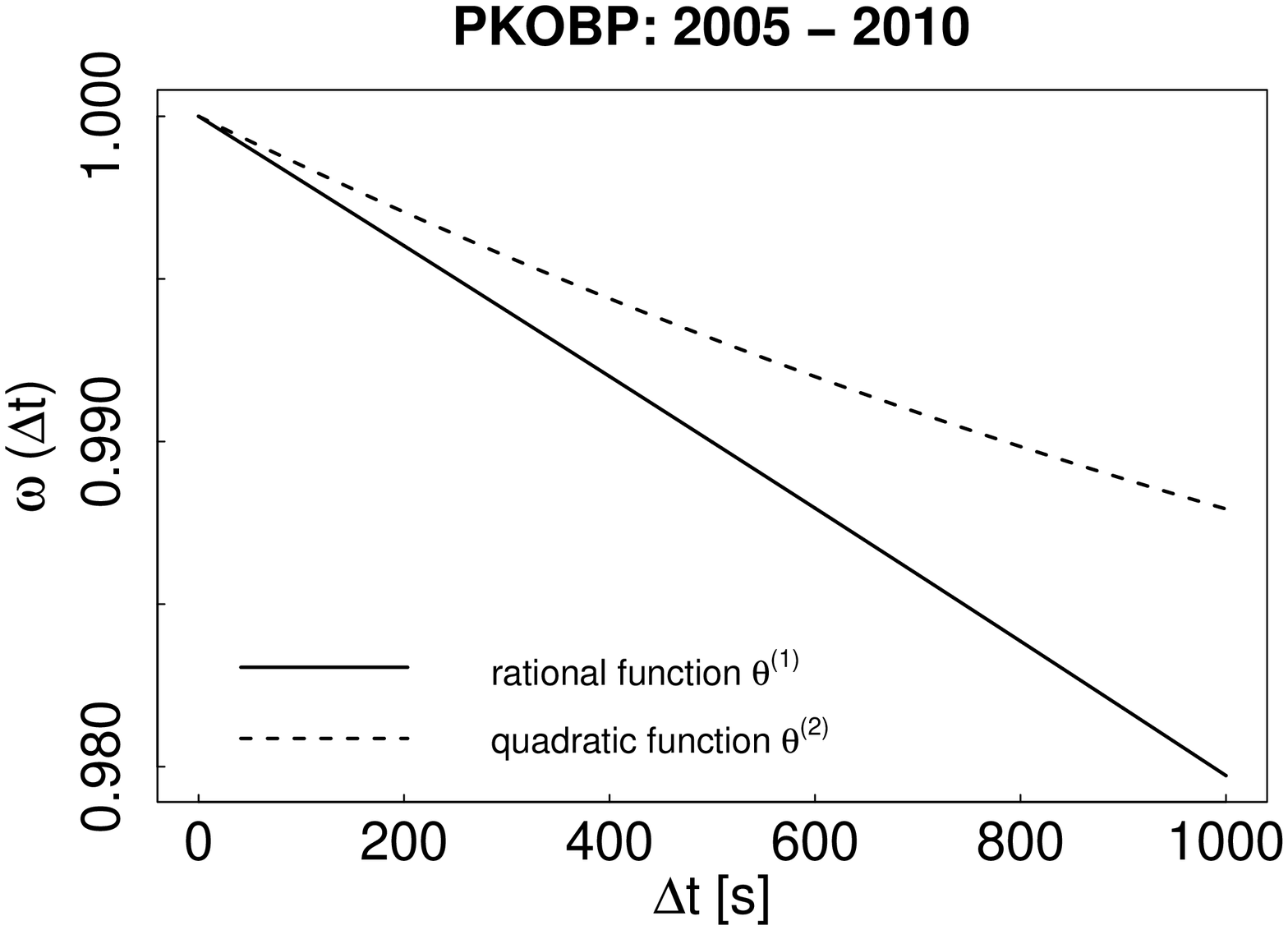}
	\end{center}
	\caption{Functions $\omega (\Delta t)$  defined in Eq. (\ref{eq:omega}) for KGHM and PKOBP, calculated by using quadratic functions $\theta^{(1)}$ (dashed curve) and rational functions $\theta^{(2)}$ (solid curve).}
	\label{pic:weight}
\end{figure}

Next, we use the monotonicity of the autocorrelation function for time lag $\Delta 1 > 0$. 
We verify whether the argument of the autocorrelation of the process $X$, in the last expression of above inequality, is smaller than $\Delta t$.
We analyze empirical values of the following function:
\begin{equation}
\label{eq:omega}
\omega (\Delta t) = \frac{1}{\Delta t} \int_{\Delta \tau_{min}}^{\Delta \tau_{max}} \Delta \tau \rho_{\Delta t} (\Delta \tau) d \Delta \tau.
\end{equation}
In cases of KGHM and PKOBP, for both $\theta^{(1)}$ and $\theta^{(2)}$, we obtain plots of $\omega (\Delta t)$ presented in Fig. \ref{pic:weight}. 
These results (and analogous analysis, made by us, for several other stocks) allow the conclusion that $\omega (\Delta t)$ is the decreasing function and equals 1 for $\Delta t = 0$. 
Hence, for increasing and concave autocorrelation functions we get:
\begin{equation}
\label{eq:ineq2}
C_Y ( \Delta t ) \leq C_X ( \Delta t ),
\end{equation}
where the equality holds for $\Delta t = 0$. 
Obviously, without loss of generality, we can normalize both autocorrelations. 
Then, we obtain $C_Y (0) = C_X (0) = 1$ and what is even more significant:
\begin{equation}
\lim_{\Delta t^{+} \rightarrow 0} C_Y ( \Delta t ) = \lim_{\Delta t^{+} \rightarrow 0} C_X ( \Delta t ).
\end{equation}

In order to formally carry out the same reasoning for the absolute value of the price returns process, we need to use Jensen's inequality for convex function. 
As mentioned above, the autocorrelation function of absolute values of returns are positive, decreasing, and convex.
As a result, we obtain an opposite inequality for autocorrelations of processes $X$ and $Y$ (the equality still holds for $\Delta t = 0$).

\section{Concluding remarks}\label{section:res}

The main result of this work is the exact relation (\ref{eq:cor3}) between autocorrelation function estimators of seasonal (non-stationary) and stationary process. 
Furthermore, we found that for financial time series, by adding seasonality of time intervals the memory of underlying process is extended.
Fortunately, it does not create any additional autocorrelations making the relaxation time to zero longer. 
Notably, all our calculations assume nothing about processes $X$ and $Y$, except stationarity of the process $X$ and the relation (\ref{eq:base}) between them. 
Such an approach can be used in a broad range of problems and not only analysis of financial data. 
Besides, it also applies to absolute values of those processes.

\begin{acknowledgments}

We would also like to thank Professor Ryszard Kutner for his scientific guidance, productive discussions and helpful remarks.
The publication was financially supported by Foundation for Polish Science.
\end{acknowledgments}

\end{document}